

\input phyzzx


\def\frac#1#2{{#1\over #2}}

\def\del{\delta}
\def\trl{\triangleleft}
\def\tl{\tilde}
\def\pt{\partial}

\overfullrule=0pt

\def\PRL{Phys.~Rev.~Lett.}
\def\CMP{Commun.~Math.~Phys.}
\def\PL{Phys.~Lett.}
\def\MPL{Mod.~Phys.~Lett}
\def\NP{Nucl.~Phys.}

\def\IJMP{Int.~J.~Mod.~Phys.}

\def\PR{Phys.~Rev.}

\def\JSP{J.~Stat.~Phys.}


\REF\SFT{
M. Kaku and K. Kikkawa \journal \PR &D10 (74) 1110;1823;\nextline
W. Siegel \journal \PL &B151 (85) 391;396;\nextline
E. Witten \journal \NP &B268 (86) 253;\nextline
H. Hata, K. Itoh, T. Kugo, H. Kunitomo and K. Ogawa \journal \PL
&B172 (86) 186;195;\nextline
A. Neveu and P. West \journal \PL &B168 (86) 192;\nextline
M. Kaku, preprint, hep-th@xxx.lanl.gov - 9403122. 
}

\REF\IK{N. Ishibashi and H. Kawai \journal \PL &B314 (93) 190.} 

\REF\KKMW{H. Kawai, N. Kawamoto, T. Mogami and Y. Watabiki \journal \PL 
&B306 (93) 19. }

\REF\FIKN{M. Fukuma, N. Ishibashi, H. Kawai and M. Ninomiya, preprint, 
YITP/K-1045, hep-th@xxx.lanl.gov - 9312175 , to appear in {\sl Nucl. Phys. B.} }

\REF\NAK{R. Nakayama \journal \PL &B325 (94) 347.}

\REF\FKN{M. Fukuma, H. Kawai and R. Nakayama  \journal \IJMP 
&A6 (91) 1385; \nextline
R. Dijkgraaf, E. Verlinde and H. Verlinde \journal \NP &B348 (91) 435.}

\REF\JR{A. Jevicki and J. Rodrigues, preprint, BROWN-HET-927, 
hep-th@xxx.lanl.gov - 9312118.}

\REF\wata{Y. Watabiki, preprint, INS-1017, hep-th@xxx.lanl.gov - 9401096.}

\REF\IKtwo{N. Ishibashi and H. Kawai \journal \PL &B322 (94) 67.}

\REF\GN{E. Gava and K.S. Narain \journal \PL &B263 (91) 213.}

\REF\FRA{M.E. Agishtein and A.A. Migdal \journal \IJMP &C1 (90) 165; 
 {\sl Nucl.~Phys.} {\bf B350} (1991), 690.}

\REF\ni{N. Ishibashi \journal \MPL &A4 (89) 251.}

\REF\cardy{J.L. Cardy \journal \NP &B324 (89) 581.}

\REF\MSS{G. Moore, N. Seiberg and M. Staudacher \journal \NP &B362 (91)
665.}

\REF\AB{J. Alfaro \journal \PR &D47 (93) 4714;\nextline
D.V. Boulatov \journal \MPL &A8 (93) 557.}

\REF\KPZ{V.G. Knizhnik, A.M. Polyakov and A.B. Zamolodchikov \journal
\MPL &A3 (88) 819; \nextline
F. David \journal \MPL &A3 (88) 1651; \nextline
J. Distler and H. Kawai \journal \NP &B321 (89) 509.}

\REF\MMS{E. Martinec, G. Moore and N. Seiberg \journal \PL &B263 (91) 190.}

\REF\DS{E. Br\'{e}zin and V. Kazakov \journal \PL &B236 (90) 144;\nextline
M. Douglas and S. Shenker \journal \NP &B335 (90) 635;\nextline
D. Gross and A. Migdal \journal \PRL &64 (90) 127; 
{\sl Nucl.~Phys.} {\bf B340} (1990), 333.}

\REF\stau{M. Staudacher \journal \PL &B305 (93) 332.}

\REF\ABF{G.E. Andrews, R.J. Baxter and P.J. Forrester \journal 
\JSP &35 (84) 193.; \nextline 
V.Pasquier \journal \NP &B285 (87) 162.}

\REF\FKNtwo{M. Fukuma, H. Kawai and R. Nakayama \journal \CMP &143 (92) 371.}

\FIG\SDPRO{Processes involved in S-D equations. If one deforms the loop on the 
left hand side at the cross, it either splits into two (the first term on the 
right hand side), absorbs another loop (the second term) or changes in its 
spin configuration (the third term). The change in the spin configuration 
is expressed by an operator ${\cal H}$.}

\FIG\H{The action of the operator ${\cal H}$.}

\FIG\HHone{The S-D equation (2.5).}

\FIG\HHtwo{The S-D equation (2.6).}

\FIG\MXCON{The mixed spin configuration.}

\FIG\HHthree{The S-D equation (3.4).}

\FIG\HHH{The S-D equations (3.5).}

\FIG\star{The product $*$.}

\titlepage

\rightline{KEK-TH-402}
\rightline{KEK preprint 94}
\rightline{EPHOU-94-003}

\title{String Field Theory in the Temporal Gauge}

\author{M.~Ikehara$^{1,2}$, N.~Ishibashi$^{1}$, H.~Kawai$^{1}$, 
T.~Mogami$^{1,3}$, R.~Nakayama$^{4}$ and N.~Sasakura$^{1}$}
\address{$^{1}$KEK theory group, Tsukuba, Ibaraki 305, Japan}
\address{$^{2}$Department of Physics, University of Tokyo,
Bunkyo-ku, Tokyo 113, Japan}
\address{$^{3}$Department of Physics, Kyoto University, 
Kitashirakawa, Kyoto 606, Japan}
\address{$^{4}$Department of Physics, Hokkaido University,
Sapporo 060, Japan}

\abstract{
We construct the string field Hamiltonian for $c=1-\frac{6}{m(m+1)}$ 
string theory in the temporal gauge. In order to do so, we first examine the 
Schwinger-Dyson equations of the matrix chain models and propose the 
continuum version of them. Results of boundary conformal field theory 
are useful in making a connection between the discrete and continuum pictures. 
The $W$ constraints are derived from the continuum Schwinger-Dyson equations. 
We  
also check that these equations are consistent with 
other known results about noncritical string theory. 
The string field Hamiltonian 
is easily obtained from the continuum Schwinger-Dyson equations. 
It looks similar to Kaku-Kikkawa's Hamiltonian and may readily be 
generalized to $c>1$ cases. 
}

\endpage


\chapter{Introduction}

String theory provides us with the most promising framework for describing the 
physics at the Planck scale. However, a 
nonperturbative treatment of string theory is indispensable for relating it 
to the lower energy phenomena we see. String field theory \refmark{\SFT} 
is expected 
to make such treatment possible. A string field theory corresponds to a rule to 
cut the string worldsheets into vertices and propagators, or in other words, a 
way to fix the reparametrization invariance. 

Recently a new class of string field theory is 
proposed for $c=0$ noncritical string \refmark{\IK}. 
It is based on a gauge fixing \refmark{\KKMW} 
of the reparametrization invariance, 
which can naturally be considered on dynamically triangulated worldsheets. The 
gauge, which can be called the temporal gauge \refmark{\FIKN } or the 
proper time gauge\refmark{\NAK}, 
is peculiar in many 
respects. For example, in this gauge, 
even a disk amplitude is expressed as a sum of infinitely many 
processes involving innumerable splitting of strings. It forms a striking 
contrast to the case of the conformal gauge. The amplitudes can be calculated 
by using the Schwinger-Dyson (S-D) equations of the string field. 
Actually the S-D equations are powerful enough to make a 
nonperturbative treatment of the $c=0$ string possible. Indeed, the Virasoro 
constraints\refmark{\FKN} can be derived from the 
S-D equation and all the results of the matrix model are reproduced. 
Conversely it was pointed out by Jevicki and Rodrigues \refmark{\JR} 
that the string field Hamiltonian can be derived from the stochastic 
quantization of the matrix model. Also in [\wata ], the string field 
Hamiltonian was deduced from the matrix model. 

Therefore if the temporal gauge string field theory is constructed for the 
critical string, it may be a useful tool to study the nonperturbative 
effects of string theory. 
In order to go from $c=0$ to the critical string, one needs 
to know a way to introduce matter degrees of freedom on the worldsheet. 
In [\IKtwo ], $c\leq 1$ string field Hamiltonian was constructed by 
changing the way of gauge fixing a little. 
However, it was not possible to derive the $W$ constraints from this 
Hamiltonian and prove that it really describes a $c\leq 1$ string theory. 

In the present work, we will propose a string field theory of 
$c=1-\frac{6}{m(m+1)}$ string in the temporal gauge such that the $W$ 
constraints are deduced from the string field S-D equation. 
Actually we start from the matrix model S-D equations, from which the $W$ 
constraints are deduced. We propose the continuum version of these 
equations. 
Since the string field S-D equations are in close 
relation with the matrix model ones, it is easy to construct the 
string field Hamiltonian once we know the continuum version of 
matrix model S-D equations. Thus the Hamiltonian we construct is 
naturally related to the $W$ constraints. 

The organization of this paper is as follows. In section 2, we first consider 
$c=\frac{1}{2}$ case as an example. After briefly explaining the relation 
between the temporal gauge string field theory and the matrix model S-D 
equations, we examine the S-D equations for the two matrix model which 
were analysed by Gava and Narain\refmark{\GN}. We propose the continuum 
version of these equations and show that the $W_3$ constraints can be 
deduced from the continuum equations. We also check if our equations are 
consistent with other known results of $c=\frac{1}{2}$ string theory. 
In section 3, we generalize the discussion of section 2 to the case of 
$c=1-\frac{6}{m(m+1)}$ string. In section 4, we construct the string 
field Hamiltonian from the S-D equations obtained in section 2 and 3.

\chapter{Continuum S-D Equations for $c=\frac{1}{2}$ String}

Let us recall the definition of the time coordinate in [\KKMW]. 
Suppose a randomly triangulated surface with boundaries. The time coordinate 
of a point on the surface is defined to be the length of the shortest curves 
connecting the point and the boundaries. 
In [\KKMW], this time coordinate was introduced to study the fractal structure
\refmark{\FRA}  
of random surfaces. 
It was shown that a well-defined 
continuum limit of such a time coordinate exists at least in the case of 
$c=0$ string. If one takes such a time coordinate $t$ 
in the continuum limit, the metric will look like 
$$
ds^2=dt^2+h(x,t)(dx+N^1(x,t)dt)^2.
$$
In [\FIKN ], 2D quantum gravity was studied by further fixing the gauge as 
$\partial_xh=0$. Such a gauge was called the temporal gauge. 
In [\NAK ], the gauge $N^1=0$ was pursued, which was called the proper time 
gauge. 

In this coordinate system, we cut the surface into time 
slices. Then the surface can be interpreted as describing a history of 
strings which keep splitting and joining. 
In [\IK], a string field Hamiltonian 
$H$ describing the 
evolution of the strings in such a coordinate system was constructed. 
In this paper, we will call this Hamiltonian the string field Hamiltonian 
in the temporal gauge. 
( It can also be called the proper time gauge Hamiltonian. )
$H$ is expressed in terms of the creation (annihilation) operator 
$\Psi ^\dagger (l)$ ($\Psi (l)$) of the string. 
Since each string is 
labelled only by its length, the string field is a function of the length $l$. 
An $n$-string amplitude corresponds to the worldsheets with $n$ boundaries, 
each of which describes an external string state. Therefore such an amplitude 
is expressed as 
$$
\lim_{D\rightarrow \infty }<0|e^{-DH}
\Psi^\dagger (l_1)\cdots \Psi^\dagger (l_n)|0>.
\eqn\stoc
$$
The string amplitudes can be obtained by solving the string field 
S-D equation:
$$
\lim_{D\rightarrow \infty }\partial_D<0|e^{-DH}
\Psi^\dagger (l_1)\cdots \Psi^\dagger (l_n)|0>=0.
\eqn\stosd
$$
This equation means that the string amplitudes do not change if one acts 
the time evolution operator on all the external string states. In the 
point of view of 2D quantum gravity, this equation corresponds to 
the Wheeler-DeWitt equation. 

Even if there are matter fields on a dynamically triangulated  surface, 
a time coordinate can be defined in the same way. Here we concentrate on 
$c=\frac{1}{2}$ string. Such a string can be realized by 
putting the Ising model on the random surface. Since the length 
of a curve on the surface 
is defined irrespective of the matter, the time coordinate can be defined 
and the surface is cut into time slices. Again, the surface can be regarded as 
describing a history of strings. Therefore we will be able to construct a 
string field Hamiltonian describing the time evolution of the strings. 
However, in this case, the strings have Ising spins on them. Hence the string 
field depends not only on the length of the string but also on the spin 
configuration on it.  

In the continuum limit, an Ising spin configuration 
may be represented by a state\foot{Here the states we mean do not necessarily 
satisfy the condition $(L_0-\bar{L}_0)|v\rangle =0$. } 
 of $c=\frac{1}{2}$ 
conformal field theory (CFT). 
The splitting and joining of the strings should be 
described by the three-Reggeon-like vertex for  $c=\frac{1}{2}$ CFT and the 
string field Hamiltonian will be very complicated. This is the reason why 
an alternative definition of the time coordinate was taken in 
[\IKtwo]. Here we would like to 
stick to this time coordinate and construct the Hamiltonian in the temporal 
gauge. 

One can obtain a hint on the form of such a Hamiltonian by examining the 
matrix model S-D equations. 
As was discussed in [\IK ,\wata ], the string field S-D 
equations are closely related to the matrix model S-D equations. 
The latter describe the change of the partition functions 
corresponding to dynamically triangulated surfaces when one takes 
a triangle away from a boundary. It is obvious from the definition of the 
time coordinate that 
at the discrete level the former equations describe the changes which happen 
when one takes one layer of triangles from all the boundaries. Therefore, in the 
continuum limit, the former should be expressed as an integration  
of the latter. 

Hence if we know the continuum limit of the matrix model S-D 
equations, we can figure out what the string field S-D equations 
should be. Then we can infer the form of the string field Hamiltonian. 
Gava and Narain \refmark{\GN} studied the S-D 
equation for the two matrix model and deduced the $W_3$ constraints. 
In this section, we will consider the continuum limit of the Gava and Narain's 
equations. 

\section{Continuum Limit of Gava-Narain's Equation}
Let us sketch how Gava and Narain obtained the $W_3$ constraints. 
The $W_3$ constraints are expected to come 
from equations about the loop amplitudes in which 
the Ising spins on all the 
boundary loops are, say, up. 
Suppose the partition function of the 
dynamically triangulated surfaces with boundaries on which all the Ising spins 
are up. If one takes one triangle from a boundary, the following three things 
can happen. (Fig.\SDPRO )

\item{1.} The boundary loop splits into two. 
\item{2.} The boundary loop absorbs another boundary. 
\item{3.} The spin configuration on the boundary loop changes. 

The matrix model S-D equation is a sum of three kinds of terms 
corresponding to the above processes. 
In the first and the second process, only boundaries with all the spins up 
can appear. The third process is due to the matrix model action. 
A boundary loop on which one spin is down and all the others are up can 
appear in this process. In order to derive the $W_3$ constraints, one should 
somehow cope with this mixed spin configuration. Gava and Narain then 
considered the loop amplitudes with one loop having such a spin configuration 
and all the other loops having all the spins up. 
They obtained two S-D equations corresponding to the processes of 
taking away the triangle attached to the link on which the Ising spin is down 
and the one attached to the next link. 
Those equations also consist of the terms corresponding to the above three 
processes. 
With these two equations, one can 
express the loop amplitude with one mixed spin loop insertion by loop amplitudes 
with all the spins up. Thus they can obtain closed equations for loop amplitudes 
with all the spins up and the $W_3$ constraints were derived from them. 

We would like to rewrite the above procedure in terms of the continuum 
language. Let us define 
the continuum loop operator $w(l;|v\rangle )$ 
as representing a loop with length $l$ 
and the spin configuration corresponding to $|v\rangle $ which is a state 
of $c=\frac{1}{2}$ CFT. We take the loop to have one marked point on it. 
The loop amplitude will be denoted by 
$$
<w(l_1;|v_1\rangle )w(l_2;|v_2\rangle )\cdots w(l_n;|v_n\rangle )>.
\eqn\ampl
$$

The matrix model S-D equation describes the change of the amplitude 
eq.\ampl , when one takes a triangle away from a boundary. Now we will 
construct the continuum version of it, which describes what happens when one 
deforms the amplitude eq.\ampl\ at a point on a boundary. 
In principle, by closely looking at the discrete S-D equations and taking the 
continuum limit, one should be able to figure out what the continuum S-D 
equations will be. However, in actuality, it is not an easy task, because of 
the existence of the non-universal parts in the loop operators and the operator 
mixing between various loop operators. Therefore, here we will construct 
the continuum S-D equations by assuming the following properties of them and 
check the validity of our assumption later by deriving the $W$ constraints 
from them.  

\item{1.} We will assume that the continuum S-D equation also consists of 
the three 
terms representing the three processes in the above (Fig.\SDPRO ), i.e. a loop 
splitting into two, a loop absorbing another one and changes in the spin 
configuration of the loop. Let us call the first two the vertex terms and 
the last one the kinetic term. 
\item{2.} We will assume that when a loop 
splits into two, the descendant loops should inherit the spin configuration 
of the original loop. Such a three string vertex will be expressed by 
a delta functional of the spin configurations, i.e. the 
three-Reggeon-like vertex 
of $c=\frac{1}{2}$ CFT in the continuum limit. 
The process where a loop absorbs another loop will 
also be expressed by the three-Reggeon-like vertex. 
\item{3.} In the matrix model S-D equations, the kinetic terms come from the 
matrix model action. In the two matrix model, they include terms which 
change the length of the loop as well as a term which flips the spin. 
We will assume that in the continuum limit, only the spin flipping term 
survives. 

With all these assumptions, we are able to write down the continuum S-D 
equations. 
We will present the 
most general continuum S-D equation using such vertices in section 4. Here 
let us concentrate on a simpler situation, which Gava and Narain considered. 
In the derivation of the $W_3$ constraints, they started from loops with 
all the spins up. Such a spin configuration was represented as a state 
of $c=\frac{1}{2}$ CFT in [\ni ,\cardy]. 
Let us denote such a state by $|+\rangle $. 
It is clear that if such a loop splits into two, it results in 
two loops with all the spins up. Also if a loop with all the spins up absorbs 
another one, we obtain another loop with all the spins up. Therefore 
the process of splitting and merging is particularly simple for such kind of 
loops. The first S-D equation Gava and Narain considered corresponds to the 
deformation of the loop amplitude eq.\ampl\ with $|v_1\rangle =|v_2\rangle 
=\cdots =|v_n\rangle =|+\rangle $. The equation in the continuum limit should 
be 
$$
\eqalign{
&\int_0^l dl'<w(l';|+\rangle )w(l-l';|+\rangle )
w(l_1;|+\rangle )\cdots w(l_n;|+\rangle )> \cr
&+g\sum_kl_k<w(l+l_k;|+\rangle )w(l_1;|+\rangle )\cdots w(l_{k-1};|+\rangle )
w(l_{k+1};|+\rangle )\cdots w(l_n;|+\rangle )>\cr
&+<w(l;{\cal H}(\sigma )|+\rangle )w(l_1;|+\rangle )\cdots w(l_n;|+\rangle )>
\approx 0.}
\eqn\eqone
$$
Here the first term corresponds to the process 1 in the above and the second 
term is for the process 2. The string coupling constant $g$ comes in front 
of the second term as in the case of $c=0$ string\refmark{\IK}. 
The last term describes the process 3, where the 
operator ${\cal H}(\sigma )$ expresses the local change of the spin 
configuration. $0\leq \sigma < 2\pi$ is the coordinate of the point where 
the local change occurs. The coordinate $\sigma $ on the loop is taken so that 
the induced metric on the loop becomes independent of $\sigma $. 
$\sigma =0$ is taken to be the marked point of the 
loop. 
$\approx 0$ here means that 
as a function of $l$, the quantity has its support at $l=0$. Therefore the 
left hand side of eq.\eqone\ is equal to a sum of derivatives of $\delta (l)$. 
These delta functions correspond to processes in which a string with  
vanishing length disappears. In the point of view of string field theory, 
such processes are expressed by the tadpole terms. 

Therefore $w(l;{\cal H}(\sigma )|+\rangle )$ 
is supposed to correspond to a loop with one spin flipped to be down because of 
${\cal H}(\sigma )$, and the rest of the spins up (Fig.\H ). In the continuum 
limit, the operator which is on the domain wall of up and down spins is 
identified to be $\phi_{2,1}$\refmark{\cardy} (Fig.\H ). Therefore 
${\cal H}(\sigma )$ may be written as $\lim_{\sigma^\prime \rightarrow \sigma}
\phi_{2,1}(\sigma^\prime )\phi_{2,1}(\sigma )$. With this operator, we can 
express everything about the S-D equations in terms of the continuum language. 

The two other equations which Gava and Narain used were obtained by taking a 
triangle away from 
$w(l;{\cal H}(\sigma )|+\rangle )$. The triangles to be considered were 
the one attached to the link where ${\cal H}(\sigma )$ is inserted and the 
one next to it. 
In the continuum limit, these equations will correspond to the following two 
equations. In one of the equations, we consider a loop 
$w(l;{\cal H}(\sigma )|+\rangle )$ and deform at a point near $\sigma $ and 
take the limit in which the point tends to $\sigma $ (Fig.\HHone ). The 
S-D equation becomes 
$$
\eqalign{
&\int_0^l dl'<w(l';|+\rangle )w(l-l';{\cal H}(\sigma )|+\rangle )
w(l_1;|+\rangle )\cdots w(l_n;|+\rangle )>\cr
&+g\sum_kl_k<w(l+l_k;{\cal H}(\sigma )|+\rangle )
w(l_1;|+\rangle )\cdots w(l_{k-1};|+\rangle )
w(l_{k+1};|+\rangle )\cdots w(l_n;|+\rangle )>
\cr
&
+<w(l;({\cal H}(\sigma ))^2|+\rangle )w(l_1;|+\rangle )\cdots w(l_n;|+\rangle )>
\approx 0.
\cr
}
\eqn\eqtwo
$$
Here $w(l;({\cal H}(\sigma ))^2|+\rangle )$ denotes the limit 
$\lim_{\sigma ' \rightarrow \sigma}w(l;{\cal H}(\sigma ')
{\cal H}(\sigma )|+\rangle )$. 
In the other equation, we consider a loop 
$w(l;\phi_{2,1}(\sigma^\prime )\phi_{2,1}(\sigma )|+\rangle )$, deform at 
a point between the two $\phi_{2,1}$ insertions and then take the limit 
$\sigma ' \rightarrow \sigma $ (Fig.\HHtwo ). The insertion of ${\cal H}$ yields 
$w(l;({\cal H}(\sigma ))^2|+\rangle )$ again. However, this time, 
the loop cannot split or absorb a loop $w(l;|+\rangle )$. When a loop splits, 
two points on the loop should merge. The spin configurations at the two 
points should coincide in order for this to happen. Now we deform the 
loop at a point in the down spin region and the point it merges with 
should be in the down spin region. Therefore, in the limit 
$\sigma ' \rightarrow \sigma $, no splitting can occur. The loop cannot 
absorb another loop for the same reason. Hence we obtain 
$$
<w(l;({\cal H}(\sigma ))^2|+\rangle )w(l_1;|+\rangle )\cdots w(l_n;|+\rangle )>
\approx 0.
\eqn\eqthree
$$
This equation means that the loop $w(l;({\cal H}(\sigma ))^2|+\rangle )$ is 
in a sense ``null''. Similar arguments as above show that correlation functions 
involving such a loop vanishes unless there exist any finite regions 
of down spins on the boundaries. 

We propose eqs.\eqone, \eqtwo\ and \eqthree\ as the continuum limit of the 
Gava-Narain's equations. As a check of the validity of our equations, let 
us first see the disk amplitude of $c=\frac{1}{2}$ string theory satisfies these 
equations. Let us denote the disk amplitude with a loop 
$w(l;({\cal H}(\sigma ))^n|+\rangle )$ as 
the boundary by $w_n(l)$\foot{Because of the reparametrization invariance, 
correlation functions involving $w(l;({\cal H}(\sigma ))^n|+\rangle )$'s 
do not depend on $\sigma $.}. 
The Laplace transform $\tl w_0(\zeta )
=\int_0^\infty dle^{-\zeta l}w_0(l)$ is known as
\refmark{\MSS},
$$
\tl w_0(\zeta )
=
(\zeta +\sqrt{\zeta^2 -t})^{\frac{4}{3}}+
(\zeta -\sqrt{\zeta^2 -t})^{\frac{4}{3}},
\eqn\disk
$$
where $t$ is the cosmological constant. If our equations really correspond to 
$c=\frac{1}{2}$ string theory, this disk amplitude should satisfy these 
equations at the lowest order in the expansion in terms of $g$. In the Laplace 
transformed form, the equations to be satisfied are,
$$
\eqalign{
&(\tl w_0(\zeta ))^2
+\tl w_1(\zeta )\approx 0,
\cr
&\tl w_0(\zeta )\tl w_1(\zeta )
+\tl w_2(\zeta )\approx 0,
\cr
&\tl w_2(\zeta )\approx 0.}
\eqn\diskeq
$$
Here $\tl w_n(\zeta )$ denotes 
$\int_0^\infty dle^{-\zeta l}w(l;{\cal H}^n(\sigma )|+\rangle )$. 
$\approx 0$ here means that 
the quantity is a polynomial of $\zeta$. 
It is easy to see that the disk amplitude \disk\ and 
$$
\tl w_1(\zeta )
=
(\zeta +\sqrt{\zeta^2 -t})^{\frac{8}{3}}+
(\zeta -\sqrt{\zeta^2 -t})^{\frac{8}{3}}-t^{\frac{4}{3}},
\eqn\diskone
$$
is a solution of \diskeq.

This $w_1$ in eq.\diskone\ is a new kind of amplitude which has never 
appeared in the literature. 
It indeed emerges in the continuum limit 
of the matrix model disk amplitude $W(P)=<\frac{1}{N}tr(P-A)^{-1}>$\foot{
Here $A$ denotes one of the matrices in the two matrix model. Here we follow the 
notation of the reference [\stau ].}. $W(P)$ is a solution to the 
matrix model S-D equations given in [\GN, \AB, \stau ]. 
In the continuum limit, one should take $P$ and the matrix model coupling 
constant $g$\foot{Don't confuse it with the string coupling constant $g$.}
to approach the critical value 
$P_\ast $  and $g_\ast$ as $P=P_\ast +a\zeta ,~g=g_\ast +{\rm const.}a^2t$, 
where $a$ is the lattice cutoff. 
By expanding $W(P)$ 
in powers of $a$, one obtains 
$$
W(P)=b_0+b_3\zeta a+b_4\tl w_0(\zeta )a^{\frac{4}{3}}
+b_5\partial_\zeta \tl w_1(\zeta )a^{\frac{5}{3}}+O(a^2),
$$
where $b_i$'s are non-universal constants. Thus we can see that not only 
$w_0(l)$ but also $lw_1(l)$ are included in the continuum limit of the disk 
amplitude $W(P)$. Here 
$lw_1(l)\sim <w(l;l\int d\sigma {\cal H}(\sigma )|+\rangle )>_0$ 
rather than $w_1(l)$ appears because  
$W(P)$ corresponds to a loop which is invariant under rotation. 

We conclude this subsection with a comment on the scaling dimensions. 
The scaling dimension of the disk amplitude $<w(l;|+\rangle )>_0$ can be 
estimated \refmark{\MSS} by the KPZ-DDK argument \refmark{\KPZ} to be 
$L^{-\frac{7}{3}}$, where $L$ denotes the dimension of the boundary length. 
>From the above result, the dimension of 
$<w(l ;{\cal H}(\sigma )|+\rangle )>_0$ is $L^{-\frac{11}{3}}$. 
The difference $L^{-\frac{4}{3}}$ of the dimensions is attributed to the 
insertion of the operator ${\cal H}(\sigma )$. 
Notice that eqs.\diskeq\ make sense as a continuum S-D equation only when 
${\cal H}(\sigma )$ has such a dimension. 
It is quite consistent with 
the identification ${\cal H}(\sigma )=\lim_{\sigma^\prime \rightarrow \sigma}
\phi_{2,1}(\sigma^\prime )\phi_{2,1}(\sigma )$, because the gravitational 
scaling dimension of $\phi_{2,1}$ on the boundary is \refmark{\KPZ,\MMS} 
estimated to be $L^{-\frac{2}{3}}$. 

\section{Derivation of the $W_3$ Constraints}
If our continuum limit S-D equation is correct, eqs.\eqone,
\eqtwo\ and \eqthree\ should yield the $W_3$ constraints. 
In this subsection we will show that 
this is indeed the case. In order to do so, let us define the generating 
functional of loop amplitudes as 
$$
\eqalign{
Z&(J_0(l),J_1(l),J_2(l))\cr
&=<\exp (\int_0^\infty dlJ_0(l)w(l;|+\rangle )
+\int_0^\infty dlJ_1(l)w(l;{\cal H}(\sigma )|+\rangle )
+\int_0^\infty dlJ_2(l)w(l;({\cal H}(\sigma ))^2|+\rangle ))>.
}
\eqn\noone
$$
Using this generating functional, the S-D equations
\eqone, \eqtwo\ and \eqthree\ can be rewritten as
$$
\eqalign{
&({\delta \over \delta J_{n+1}(l)}
+\int_0^ldl' {\delta^2 \over \delta J_0 (l') \delta J_n(l-l') }
+\int_0^\infty dl'l'J_0(l'){\delta\over \delta J_n(l+l')})
Z|_{J_i(l)=0\ (i=1,2)}\approx 0\ (n=0,1),\cr
&{\delta \over \delta J_2(l)}Z|_{J_i(l)=0\ (i=1,2)}\approx 0.}
\eqn\sdeqz
$$
Here we have set the string coupling $g=1$ for notational simplicity. 
The fact that the left hand side of the three equations above do not 
vanish unless $l\neq 0$ makes further analysis cumbersome. 
We can see from the analysis of the disk amplitudes in the above 
that the tadpole terms should exist. 
However it is possible to show that we can cancel such tadpole term 
contributions by shifting $J_0(l)$ as 
$J_0(l)\rightarrow c_1l^{\frac{1}{3}}+c_2l^{\frac{7}{3}}+J_0(l)$, 
and we obtain the equations 
\sdeqz\ with $\approx $ replaced by $=$. Indeed the $W$ constraints are 
usually written in terms of such shifted variables\refmark{\FKN}. For 
notational simplicity, we will deal with equations which are obtained 
after such a shift is done. 


It is convenient to use the notations
$$
\eqalign{
(f\ast g)(l)&=\int_0^l dl' f(l-l')g(l'),\cr
(f\trl g)(l)&=\int_0^\infty dl' f(l')g(l+l'). 
}
\eqn\asttrl
$$
Then the first line of \sdeqz\ 
can be rewritten in a simpler form: 
$$
({\del \over \del J_{n+1}(l)}
+({\del \over \del J_0}\ast {\del \over \del J_n})(l) 
+((lJ_0)\trl {\del \over \del J_n})(l)
)Z|_{J_i(l)=0\ (i=1,2)}=0\ (n=0,1).
\eqn\dfsd
$$ 
By solving $\del /\del J_2(l)$ in terms of 
$\del / \del J(l)$ and $lJ(l)$, and substituting it into
the second line of \sdeqz, we obtain
$$
(({\del \over \del J_0}\ast +(lJ_0) \trl)^2
{\del \over \del J_0})(l) Z=0.
\eqn\sdm
$$
Here $J_i(l)=0\ (i=1,2)$ is implicitly understood. 
Also we always understand $\trl$ as an operation to the right:
$A_1\trl A_2 \trl \cdots \trl A_n =
(A_1\trl(A_2\trl(\cdots \trl A_n)\cdots )$.
To deduce the $W_3$ constraints from \sdm, one should subtract
the non-universal part of $Z$. 
In usual, the non-universal parts exist in the disk and
the cylinder amplitudes.
However after the shift of $J_0(l)$ discussed in the above, the disk 
amplitude vanishes. 
Hence only the subtraction of the non-universal part of the 
cylinder amplitude is needed:
$$
\eqalign{
Z&=Z_{non}Z_{univ.},\cr
Z_{non}&=\exp({1\over 2}\int_0^\infty 
dl dl' J_0(l) C_{non}(l,l') J_0(l')),
}
\eqn\sepz
$$
Then, substituting \sepz\ into \sdm, we obtain the S-D
equation for 
the universal part of the partition function merely by shifting 
the derivative:
$$
{\del \over \del J_0(l)}=D(l)+\int_0^\infty
dl' C_{non}(l,l') J_0(l'),
\eqn\sdmd
$$
where $D(l)$ denotes the derivative for the universal part.

Next we will specify the $C_{non}(l,l')$ 
and $D(l)$, and deduce the $W_3$ constraints explicitly.
It is more convenient to
work in the Laplace transformed variables:
$$
\tl f(\zeta)=\int_0^\infty dl \exp(-l\zeta) f(l).
\eqn\notwo
$$
In such variables, the operations $\ast$ and $\trl$ defined
in \asttrl\ are expressed as
$$
\eqalign{
(\tl f \ast \tl g)(\zeta)&=\tl f(\zeta)\tl g(\zeta),\cr
(\tl f \trl \tl g)(\zeta)&=-\int {d\zeta'\over 2\pi i} \tl f(-\zeta')
{\tl g(\zeta)-\tl g(\zeta') \over \zeta-\zeta'}.
}
\eqn\nothree
$$

The non-universal part can be obtained by the orthogonal polynomial technique 
\refmark{\DS ,\MSS}. 
Substituting it into \sdmd, we obtain
$$
{\del \over \del \tl J_0(\zeta)}=\tl D(\zeta)
+{1\over 3}\int {d\zeta'\over 2\pi i} \tl K(-\zeta')
{2-({\zeta'\over \zeta})^{1\over 3} 
-({\zeta'\over \zeta})^{2\over 3}
\over \zeta-\zeta'}, 
\eqn\nonunitwo
$$
where $\tl K(\zeta)=-\tl J^\prime_0(\zeta)=\int dl \exp (-l\zeta)lJ_0(l)$.
The universal part of the partition function depends only on
some fractional moments of the currents
$\tl J_r=\int d\zeta \tl J_0(-\zeta) \zeta^{-r-1} 
(r=n+{1\over 3},n+{2\over 3})$ with $n$ non-negative integers. 
So the $D(\zeta)$ will be expanded in the following form:
$$
\eqalign{
\tl D(\zeta)&=\sum_{r>0} \zeta^{-r-1} {\pt \over \pt \tl J_r}, \cr
r&=n+{1\over 3},n+{2\over 3}\cr}
\eqn\unider
$$
with $n$ non-negative integers.

Substituting \nonunitwo\ and \unider\ into \sdm, we have obtained
the following result after a long calculation:
$$
(:\lbrack (\tl D+{\bar K \over 3})^3 \rbrack_{\le-1} : 
+\frac{3}{2}(\tl D+{1\over 3}\hat K_3)
(:\lbrack (\tl D+{\bar K \over 3})^2 \rbrack_{\le-1} : 
+{2\over 27\zeta^2}))Z_{univ.}=0,
\eqn\final
$$
where 
$$
\eqalign{
\bar K&=\sum_{r>0} \zeta^{r-1} r \tl J_r, \cr
\hat K_3 \tl f
&={1\over 2\pi i}\int d\zeta' {\tl K(-\zeta') \over \zeta-\zeta'} 
(2\tl f(\zeta')
-(({\zeta'\over \zeta})^{1\over 3}+({\zeta'\over \zeta})^{2\over 3})
\tl f(\zeta)),\cr
}
\eqn\reone
$$
and $\lbrack \cdot \rbrack_{\le-1}$ 
means taking all the terms with negative integral powers of
$\zeta$, and $::$ denotes the normal ordering such that 
$\pt/\pt \tl J_r$'s are put on the right of $\tl J_r$'s.


Expanding eq.\final\ asymptotically in powers of $\zeta^{-1}$, 
we obtain the following constraints for the partition functions:
$$
\eqalign{
W^{3}_n Z_{univ.}&=0\ \ (n=-2,-1,\cdots),\cr
L_n Z_{univ.}&=0\ \ (n=-1,0,\cdots),\cr
}
\eqn\nofour
$$
where $L_n$ and $W^3_n$ are defined through
expanding in $\zeta$ the operators appearing in \final:
$$
\eqalign{
\sum_{n=-2}^\infty W^{3}_n \zeta^{-n-3}
&=:\lbrack (\tl D+{\bar K \over 3})^3 \rbrack_{\le-1} :,\cr
\frac{2}{3}\sum_{n=-1}^\infty L_n \zeta^{-n-2}
&=:\lbrack (\tl D+{\bar K \over 3})^2 \rbrack_{\le-1} : 
+{2\over 27\zeta^2}.\cr
}
\eqn\nofive
$$
\nofour\ coincides with the $W_3$ constraints\refmark{\FKN} for the partition
function.

\section{Loop with Mixed Spin Configurations}
So far in this section, we have mainly dealt with only loops with all 
the spins up. 
As a check of the validity of our continuum 
S-D equations, we will show that they can be applied to the loops 
with mixed spin configurations which was considered in [\stau ]. 

Let us consider a loop which is divided into two connected regions of up and 
down spins. We denote such a loop by $w(l_1,l_2)$ where $l_1$ and $l_2$ 
are the length of the up and down regions respectively (Fig.\MXCON ). 
We will discuss the disk amplitude $<w(l_1,l_2)>_0$ 
with such a boundary in this subsection.  
In [\stau ], the discrete counterpart $W^{(2)}(P,Q)$ of 
$$
\tl w(\zeta_1,\zeta_2)
=\int_0^\infty dl_1\int_0^\infty dl_2e^{-\zeta_1 l_1-\zeta_2 l_2}
<w(l_1,l_2)>_0
$$
was given by solving the matrix model S-D equations. By taking the continuum 
limit of $W^{(2)}(P,Q)$ one can obtain $\tl w(\zeta_1,\zeta_2)$. 
It turns out that $\tl w(\zeta_1,\zeta_2)$ mixes with $\tl w_0(\zeta_1)$ and 
$\tl w_0(\zeta_2)$. Therefore we should subtract a multiple of $W(P)+W(Q)$ from 
$W^{(2)}(P,Q)$ in taking the continuum limit. In the continuum limit, 
$a\rightarrow 0,~P=P_\ast +a\zeta_1,~Q=P_\ast +a\zeta_2$, one obtains the 
expansion 
$$
W^{(2)}(P,Q)-\alpha [W(P)+W(Q)]
=d_0+d_3(\zeta_1+\zeta_2)a+d_5a^{\frac{5}{3}}\tl w(\zeta_1,\zeta_2)+O(a^2),
$$
where $\alpha$ and $d_i$'s are non-universal constants. The coefficient of 
$a^{\frac{5}{3}}$ may be identified as $\tl w(\zeta_1,\zeta_2)$ which is 
given as 
$$
-\frac{\tilde{w}_0(\zeta_1)^2+\tilde{w}_0(\zeta_1)\tilde{w}_0(\zeta_2)
+\tilde{w}_0(\zeta_2)^2-3t^{\frac{4}{3}}}{\zeta_1+\zeta_2}.
\eqn\nakayama
$$

On the other hand, we can construct the continuum S-D equation for 
$<w(l_1,l_2)>_0$ as in the previous section. If we deform the loop at a point 
in the up spin region, we obtain 
$$
\int_0^{l_1^\prime }dl w_0(l)<w(l_1-l,l_2)>_0+
\int_0^{l_1^{\prime \prime}}dl w_0(l)<w(l_1-l,l_2)>_0+
<w_1(l_1,l_2)>_0
\approx 
0.
$$
Here $l_1^\prime $ and $l_1^{\prime \prime}=l_1-l_1^\prime $ 
are the distances from the point of deformation to the two domain walls 
(Fig.\MXCON ). 
$w_1(l_1,l_2)$ denotes the loop with one ${\cal H}$ insertion at the point. 
If we deform the loop $w_1(l_1,l_2)$, we obtain the 
following two equations as in the previous section: 
$$
\eqalign{
&
\int_0^{l_1^\prime }dl w_1(l)<w(l_1-l,l_2)>_0+
\int_0^{l_1^{\prime \prime}}dl w_0(l)<w_1(l_1-l,l_2)>_0+
<w_2(l_1,l_2)>_0
\approx 
0,
\cr
&
\int_0^{l_2}dl<w(l_1^\prime ,l)>_0<w(l_1^{\prime \prime },l_2-l)>_0
+<w_2(l_1,l_2)>_0
\approx 
0,}
$$
where $w_2(l_1,l_2)$ denotes the loop with two ${\cal H}$ insertions at the 
point. Since the loop $<w(l_1,l_2)>_0$ now has the down spin region, 
the second equation does not imply $<w_2(l_1,l_2)>_0\approx 0$ contrary to the 
previous case. 
By eliminating $w_1(l_1,l_2)$ and $w_2(l_1,l_2)$ from the above 
equations, one obtains a closed equation for $<w(l_1,l_2)>_0$. It is easy 
to check that the disk amplitude eq.\nakayama\ satisfies this equation. 
Although we have not tried yet, it is in principle possible to do the same thing 
for loops with more complicated spin configurations.

\chapter{Continuum S-D Equations for $c=1-\frac{6}{m(m+1)}$ String}
It is straightforward to construct continuum S-D equations for 
$c=1-\frac{6}{m(m+1)}$ string in the same way as in the previous section. 
In this section, we will elucidate $m=4$ case as an example. 
We will show that we 
can derive the $W$ constraints from the S-D equation. 

\section{S-D Equations}
As a generalization of the two matrix model, $c=1-\frac{6}{m(m+1)}$ string 
can be realized by the $(m-1)-$matrix chain model. 
The matrices $M_i$ are labelled by an integer $i~(i=1\cdots m-1)$. 
The matter degrees of freedom is represented by this ``spin'' variable $i$. 
Each spin 
can be considered to correspond to a vertex of the Dynkin diagram of 
$A_{m-1}$ so that the matrix chain potential $\sum_itr(M_iM_{i+1})$ is 
written as $\sum_{i,j}C_{ij}tr(M_iM_j)$ by the connectivity matrix $C_{ij}$ 
of the Dynkin diagram. 

In this case, a string is labelled by its length and the spin configuration. 
In the continuum limit, 
the matter configuration can be expressed by a state  
in $c=1-\frac{6}{m(m+1)}$ CFT. The $W$ constraints can be obtained by 
considering S-D equations involving strings on which all the spins are $1$. 
In [\cardy], various boundary configurations in the $A_m$ RSOS models
[\ABF] are 
identified with a state in $c=1-\frac{6}{m(m+1)}$ CFT. The RSOS realization 
of $c=1-\frac{6}{m(m+1)}$ CFT is a bit different from the matter realization 
in the matrix chain model, in which $A_{m-1}$ Dynkin diagram is related to 
$c=1-\frac{6}{m(m+1)}$. However, as in the Ising case, the fixed boundary 
conditions in the matrix chain may be identified with a boundary condition 
in which the spins on the boundary and those of the neighbors of the boundary 
are fixed in the RSOS model. Such a boundary condition is labelled by 
an integer $r~(r=1\cdots m-1)$\refmark{\cardy} 
and we will identify it with the spin 
configuration where all the spins are $r$ in the matrix chain. We will 
denote the loop on which all the spins are $1$ by $w(l;|1\rangle )$. 

The S-D equations are constructed as in the Ising case. We will illustrate 
$m=4$ case as an example. Let us consider the S-D equation corresponding to 
the deformation of a loop amplitude, 
$$
<w(l;|1\rangle )w(l_1;|1\rangle )\cdots w(l_2;|1\rangle )>.
\eqn\ampm
$$
The continuum S-D equations are constructed assuming 

\item{1.} The S-D equations consist of three kind of terms illustrated in 
Fig.\SDPRO .
\item{2.} The splitting and merging process is written by using the 
three-Reggeon-like vertex which represents a delta functional of 
the spin configurations. 
\item{3.} For the kinetic terms, only the terms in which spins are flipped 
survive in the continuum limit of the matrix model S-D equation. In the 
matrix chain model, such terms come from the matrix chain potential 
$\sum_itr(M_iM_{i+1})$. Therefore a spin $i,~1<i<m-1$ is flipped to 
$i-1$ and $i+1$, and $1$ and $m-1$ are flipped to $2$ and $m-2$ respectively. 

The equation corresponding to the deformation of eq.\ampm\ at a point on a 
boundary becomes, 
$$
\eqalign{
&\int_0^l dl'<w(l';|1\rangle )w(l-l';|1\rangle )
w(l_1;|1\rangle )\cdots w(l_n;|1\rangle )> \cr
&+g\sum_kl_k<w(l+l_k;|1\rangle )w(l_1;|1\rangle )\cdots w(l_{k-1};|1\rangle )
w(l_{k+1};|1\rangle )\cdots w(l_n;|1\rangle )>\cr
&+<w(l;{\cal H}(\sigma )|1\rangle )w(l_1;|1\rangle )\cdots w(l_n;|1\rangle )>
\approx 0.}
\eqn\eqonem
$$
${\cal H}(\sigma )$ here represents an insertion of a tiny region on which 
the spins take the value $2$. This insertion comes from the matrix chain 
potential $\sum_itr(M_iM_{i+1})$. In the continuum, the operator which is 
at the domain wall between the regions of spin $1$ and $2$ is again 
identified to be $\phi_{2,1}$\refmark{\cardy}. Therefore ${\cal H}(\sigma )$ 
insertion here can be replaced by 
$\lim_{\sigma^\prime \rightarrow \sigma }
\phi_{2,1}(\sigma^\prime )\phi_{2,1}(\sigma )$. 

We can go on to obtain equations involving 
$w(l;({\cal H}(\sigma ))^2|1\rangle )$. If one deforms 
$w(l;{\cal H}(\sigma )|1\rangle )$ at a point near $\sigma$ and take the limit 
in which the point tends to $\sigma$, one obtains
$$
\eqalign{
&\int_0^l dl'<w(l';|1\rangle )w(l-l';{\cal H}(\sigma )|1\rangle )
w(l_1;|1\rangle )\cdots w(l_n;|1\rangle )>\cr
&+g\sum_kl_k<w(l+l_k;{\cal H}(\sigma )|1\rangle )
w(l_1;|1\rangle )\cdots w(l_{k-1};|1\rangle )
w(l_{k+1};|1\rangle )\cdots w(l_n;|1\rangle )>
\cr
&
+<w(l;({\cal H}(\sigma ))^2|1\rangle )w(l_1;|1\rangle )\cdots w(l_n;|1\rangle )>
\approx 0.
\cr
}
\eqn\eqtwom
$$

So far the equations \eqonem\ and \eqtwom\ have the same form as the Ising 
case eqs.\eqone, \eqtwo. A difference comes in when one tries to obtain  
eq.\eqthree. If one deforms 
$w(l;\phi_{2,1}(\sigma^\prime )\phi_{2,1}(\sigma )|1\rangle )$ 
at a point between the two $\phi_{2,1}$ insertions and then takes the limit 
$\sigma^\prime \rightarrow \sigma $, one obtains not only the loop 
$w(l;{\cal H}^2(\sigma )|1\rangle )$ but also a loop with an insertion of 
a tiny region on which the spins are $3$ (Fig.\HHthree ). The boundary operator 
which is at the domain wall between $1$ and $3$ regions is identified 
with $\phi_{3,1}$\refmark{\cardy}. Therefore we obtain an equation 
$$
<w(l;({\cal H}(\sigma ))^2|1\rangle )w(l_1;|1\rangle )\cdots w(l_n;|1\rangle )>
+
<w(l;(\phi_{3,1}(\sigma ))^2|1\rangle )w(l_1;|1\rangle )\cdots w(l_n;|1\rangle )>
\approx 0.
\eqn\eqthreem
$$
This equation reflects the fusion rule 
$\phi_{2,1}\phi_{2,1}\sim \phi_{1,1}+\phi_{3,1}$. ${\cal H}$ should be 
identified with the $\phi_{1,1}$ part of the product $\phi_{2,1}\phi_{2,1}$. 

Thus $w(l;({\cal H}(\sigma ))^2|1\rangle )$ is not null in this case. Rather 
we can prove $w(l;({\cal H}(\sigma ))^3|1\rangle )$, which is defined as 
a limit 
$$
\lim_{\sigma_3\rightarrow \sigma_1}
w(l;{\cal H}(\sigma_3 ){\cal H}(\sigma_2 ){\cal H}(\sigma_1 )|1\rangle ),~
\sigma_3 >\sigma_2 >\sigma_1,
$$ 
is null by the following sequence of S-D equations:
$$
\eqalign{
&\lim_{\sigma_3\rightarrow \sigma_1}(
<w(l;{\cal H}(\sigma_3 ){\cal H}(\sigma_2 ){\cal H}(\sigma_1 )|1\rangle )
w(l_1;|1\rangle )\cdots w(l_n;|1\rangle )>
\cr
&~~~~~~~~~~
+
<w(l;\phi_{3,1}(\sigma_3 )\phi_{3,1}(\sigma_2 ){\cal H}(\sigma_1 )|1\rangle )
w(l_1;|1\rangle )\cdots w(l_n;|1\rangle )>)
\approx 0,
\cr
&\lim_{\sigma_3\rightarrow \sigma_1}(
<w(l;\phi_{3,1}(\sigma_3 ){\cal H}(\sigma_2 )\phi_{3,1}(\sigma_1 )|1\rangle )
w(l_1;|1\rangle )\cdots w(l_n;|1\rangle )>
\cr
&~~~~~~~~~~
+
<w(l;\phi_{3,1}(\sigma_3 )\phi_{3,1}(\sigma_2 ){\cal H}(\sigma_1 )|1\rangle )
w(l_1;|1\rangle )\cdots w(l_n;|1\rangle )>)
\approx 0,
\cr
&\lim_{\sigma_3\rightarrow \sigma_1}(
<w(l;\phi_{3,1}(\sigma_3 ){\cal H}(\sigma_2 )\phi_{3,1}(\sigma_1 )|1\rangle )
w(l_1;|1\rangle )\cdots w(l_n;|1\rangle )>)
\approx 0.}
\eqn\eqfourm
$$
Here $\sigma_3 >\sigma_2 >\sigma_1$ in all the equations. 
For example, the first equation corresponds to the deformation of the 
amplitude 
$<w(l;\phi_{2,1}(\sigma_3 )\phi_{2,1}(\sigma_2 ){\cal H}(\sigma_1 )|1\rangle )
w(l_1;|1\rangle )\cdots w(l_n;|1\rangle )>$ at a point between the two 
$\phi_{2,1}$ insertions (Fig.\HHH a). 
In the limit $\sigma_3\rightarrow \sigma_1$, 
splitting and absorbing of loops does not contribute to the equation and 
we obtain the first equation in the above. 
The derivations of the other two equations are also 
illustrated in Fig.\HHH . Thus we can prove 
$$
\lim_{\sigma_3\rightarrow \sigma_1}(
<w(l;{\cal H}(\sigma_3 ){\cal H}(\sigma_2 ){\cal H}(\sigma_1 )|1\rangle )
w(l_1;|1\rangle )\cdots w(l_n;|1\rangle )>)\approx 0.
$$

For general $m$, we can again identify ${\cal H}$ with the $\phi_{1,1}$ part 
of the product $\phi_{2,1}\phi_{2,1}$. 
We can prove by similar manipulations, 
$$
\lim_{\sigma_{m-1}\rightarrow \sigma_1}(
<w(l;{\cal H}(\sigma_{m-1})\cdots {\cal H}(\sigma_1 )|1\rangle )
w(l_1;|1\rangle )\cdots w(l_n;|1\rangle )>)
\approx 0,~(\sigma_{m-1}>\cdots >\sigma_1).
\eqn\LG 
$$
Therefore $w(l;({\cal H}(\sigma ))^{m-1}|1\rangle )$ becomes null for 
$c=1-\frac{6}{m(m+1)}$ string theory. As a generalization of eq.\eqtwom, 
we have
$$
\eqalign{
&\int_0^l dl'<w(l';|1\rangle )w(l-l';({\cal H}(\sigma ))^j|1\rangle )
w(l_1;|1\rangle )\cdots w(l_n;|1\rangle )>\cr
&+g\sum_kl_k<w(l+l_k;({\cal H}(\sigma ))^j|1\rangle )
w(l_1;|1\rangle )\cdots w(l_{k-1};|1\rangle )
w(l_{k+1};|1\rangle )\cdots w(l_n;|1\rangle )>
\cr
&
+<w(l;({\cal H}(\sigma ))^{j+1}|1\rangle )
w(l_1;|1\rangle )\cdots w(l_n;|1\rangle )>
\approx 0,
}
\eqn\eqtwog
$$
for $j=0,\cdots ,m-2$. With eqs.\LG\ and \eqtwog, the $W$ constraints will 
be derived in the next subsection. 

We will conclude this subsection with a comment on the scaling dimensions 
again. For general $m$, the scaling dimension of 
the disk amplitude $<w(l;|1\rangle )>_0$ is $L^{-\frac{2m+1}{m}}$. 
The gravitational scaling dimension of $\phi_{r,1}$ on the boundary is 
$L^{-\frac{(m+1)(r-1)}{2m}}$ and again has the right dimension for the 
continuum S-D equations to make sense. 

\section{Derivation of the W Constraints}
Let us rewrite eqs.\LG, \eqtwog\ into equations for the generating functional 
of the loop amplitudes
$$
Z^{(m)}(J_i(l))
=
<\exp (\sum_{i=0}^{m-1}
\int_0^\infty dlJ_i(l)w(l;({\cal H}(\sigma ))^i|1\rangle ))>.
$$
Eqs.\LG, \eqtwog\ become as follows:
$$
\eqalign{
&({\del \over \del \tl J_{n+1}}
+({\del \over \del \tl J_0}\ast {\del \over \del \tl J_n}) 
+\tl K\trl {\del \over \del \tl J_n})(\zeta)
Z^{(m)}|_{\tl J_i(\zeta)=0(i>0)}=0 \ (n=0,1,\cdots,m-2),\cr
&{\delta \over \delta \tl J_{m-1}(\zeta)}Z^{(m)}|_{\tl J_i(\zeta)=0
(i>0)}=0.}
\eqn\sdeqge
$$
We have assumed that the tadpole term is cancelled by an appropriate shift of 
$J_0(l)$. 
Solving $\delta/\delta \tl J_i(i>0)$'s recursively and 
substituting $\del/\del \tl J_{m-1}$ into the second line of 
\sdeqge, we obtain
$$
(({\del \over \del \tl J_0}\ast +\tl K \trl)^{m-1}
{\del \over \del \tl J_0})(\zeta) Z^{(m)}=0.
\eqn\sdmge
$$
Here $\tl J_i(\zeta)=0\ (i>0)$ is implicitly understood.
The subtraction of the non-universal part will be
$$
\eqalign{
{\del \over \del \tl J_0 (\zeta)}&=\tl D(\zeta)
+\int {d\zeta'\over 2\pi i} \tl K(-\zeta')\tl G^{(m)}(\zeta,\zeta'),\cr
\tilde G^{(m)}(\zeta,\zeta')&=
{1\over m}{m-1-\sum_{i=1}^{m-1}({\zeta'\over \zeta})^{i\over m} 
\over \zeta-\zeta'}. \cr
}
\eqn\nonunige
$$
This is a simple generalization of the known cases $m=2,3$. 


The $\tl D(\zeta)$ will be generalized to
$$
\eqalign{
\tl D(\zeta)&=\sum_{r>0} \zeta^{-r-1} {\pt \over \pt \tl J_r}, \cr
r&=n+{1\over m},n+{2\over m},\cdots,n+{m-1 \over m}
}
\eqn\uniderge
$$
with $n$ non-negative integers.

Our expectation is that, substituting \nonunige\ and \uniderge\ 
into \sdmge, one will obtain the $W_m$ constraints for the universal
part of the partition function.
We have performed the calculations explicitly for the 
cases up to $m=4$. For $m=4$, 
We have obtained, after a long calculation,
$$
\eqalign{
&(W_4(\zeta)-\frac{3}{4}[\tl KW_3]_{\ge0}(\zeta)
+{3\over 8}[\tl K[\tl KL]_{\ge0}]_{\ge0}(\zeta)
+{4\over 3}(\tl D(\zeta)+{1\over 4}K_q(\zeta))\{{3\over 4}W_3(\zeta)-{3\over 8}
[\tl KL]_{\ge0}(\zeta)\}\cr
&+{1\over 2}(2:[(\tl D(\zeta)+{1\over 4}K_q(\zeta))^2]_n:
-[\tl K\tl D]_{\ge0}(\zeta)
-{1\over4}[\tl KK_q]_{\ge0}(\zeta)
+{3\over 10}{\pt^2 \over \pt \zeta^2})L(\zeta))Z_{univ.}^{(m=4)}=0,\cr
}
\eqn\nosix
$$
where $[\cdot]_n$ means taking all the terms with non-integral powers
of $\zeta$, and 
$$
\eqalign{
[AB]_{\ge0}(\zeta)
&=-\int {d\zeta_1 \over 2\pi i} {A(-\zeta_1)B(\zeta_1)\over \zeta-\zeta_1},\cr
{1\over 2}L(\zeta)&=:[(\tl D+{1\over 4}\bar K)^2]_{\leq-1}:
+{5\over 64\zeta^2},\cr
{3\over 4}W_3(\zeta)&=:[(\tl D+{1\over 4}\bar K)^3]_{\leq-1}:,\cr
W_4(\zeta)&=:[(\tl D+{1\over4}\bar K)^4]_{\leq-1}:
-:[({\pt\over\pt\zeta}(\tl D+{1\over4}\bar K))^2]_{\leq-1}:\cr
&+({1\over 5}{\pt^2\over\pt\zeta^2}
+{15\over32\zeta^2}):[(\tl D+{1\over 4}\bar K)^2]_{\leq1}:
+{105\over(64)^2\zeta^4},
\cr
K_q(\zeta)&=
\int \frac{d\zeta_1}{2\pi i}\tl K (-\zeta_1)
\frac{(\frac{\zeta_1}{\zeta })^{\frac{1}{4}}+
(\frac{\zeta_1}{\zeta })^{\frac{1}{2}}+(\frac{\zeta_1}{\zeta })^{\frac{3}{4}}}
{\zeta -\zeta_1}.
\cr
}
\eqn\notenone
$$
Here the definition of $\bar K$ follows that in \reone\ with
the summation over $r$ following \uniderge.
Expanding eq.\nosix\ asymptotically in $\zeta^{-1}$, one obtains 
the following $W_4$ constraints for the partition functions:
$$
\eqalign{
L_n
Z_{univ.}
^{(m=4)}
&=0\ (n=-1,0,\cdots) ,\cr
W_n^3
Z_{univ.}
^{(m=4)}
&=0\ (n=-2,-1,\cdots) ,\cr
W_n^4
Z_{univ.}
^{(m=4)}
&=0\ (n=-3,-2,\cdots) ,\cr
}
\eqn\noseven
$$
where $L$'s and $W$'s are defined through expanding in $\zeta$
the operators appearing in \notenone:
$$
\eqalign{
L(\zeta)&=\sum_{n=-1} L
_n \zeta^{-n-2},\cr
W_3(\zeta)&=\sum_{n=-2} W_n^3
\zeta^{-n-3},\cr
W_4(\zeta)&=\sum_{n=-3} W_n^4
\zeta^{-n-4}.\cr
}
\eqn\noeight
$$
These coincide with the $W_4$ constraints\refmark{\FKN, \FKNtwo}. 
We conjecture that $W_m$ constraints can be derived from eqs.\sdeqge\ 
also for $m\geq 5$. 

\chapter{String Field Hamiltonian}
The discussions in the previous sections imply that the continuum S-D 
equations we proposed really describe $c=1-\frac{6}{m(m+1)}$ string theory. 
In this section we will infer the form of the string field Hamiltonian from 
these equations. 

In order to do so, we need S-D equation corresponding to the 
deformation of loops more general than $w(l;|+\rangle )$, 
$w(l;{\cal H}(\sigma )|+\rangle )$, $w(l_1,l_2)$, etc., which were 
discussed in the previous sections. 
For those loops, the vertex terms look particularly simple. In order to 
write down the continuum S-D equations for more general loops, we should 
introduce three-Reggeon-like vertex for $c=1-\frac{6}{m(m+1)}$ CFT. 
Here let us express a state of a string (with a marked point) 
as $|v\rangle_l$ by its length $l$ and 
the spin configuration $|v\rangle $. 
We define a product $*$ so that 
$$
|v_1\rangle_{l_1}*|v_2\rangle_{l_2},
$$
represents a loop made by merging the two loops $|v_1\rangle_{l_1}$ and 
$|v_2\rangle_{l_2}$ at the marked points, with the spin 
configuration inherited from them (Fig.\star ). Then the continuum S-D equation 
for generic loops will be expressed as 
$$
\eqalign{
\int_0^l dl'
&\sum_{|v^\prime \rangle ,|v^{\prime \prime }\rangle ,
~|v^\prime \rangle_{l'}*|v^{\prime \prime }\rangle_{l-l'}=|v\rangle_l}
<w(l';|v^\prime \rangle )w(l-l';|v^{\prime \prime }\rangle )
w(l_1;|v_1\rangle )\cdots w(l_n;|v_n\rangle )> \cr
&+g
\sum_kl_k\int_0^{2\pi }d\sigma^\prime
<w(l+l_k;|v\rangle_l *(e^{i\sigma^\prime {\cal P}}|v_k\rangle_{l_k}))
\cr
&~~~~~~\times
w(l_1;|v_1\rangle )\cdots 
w(l_{k-1};|v_{k-1}\rangle )
w(l_{k+1};|v_{k+1}\rangle )\cdots w(l_n;|v_n\rangle )>\cr
&+<w(l;{\cal H}(\sigma )|v\rangle )w(l_1;|v_1\rangle )\cdots w(l_n;|v_n\rangle )>
\approx 0.}
\eqn\sdeq
$$
Here ${\cal P}$ is the operator of rotation of a loop. ${\cal H}(\sigma )$ is 
identified with $\lim_{\sigma^\prime \rightarrow \sigma}
\phi_{2,1}(\sigma^\prime )\phi_{2,1}(\sigma )$. 

The S-D equation describes a deformation of a loop at a point on 
it. If we integrate it over the position of the point, we obtain the deformation 
induced by the string field Hamiltonian in the temporal gauge. 
Let $\Psi (l;|v\rangle )$ 
($\Psi^\dagger (l;|v\rangle )$) denotes the annihilation (creation) operator 
of a string with length $l$ and the spin configuration $|v\rangle $ satisfying 
$$
[\Psi (l;|v\rangle ), \Psi^\dagger (l';|v'\rangle )]
=
l\int_0^{2\pi} d\sigma \langle v'|e^{i\sigma {\cal P}}|v\rangle \delta (l-l').
\eqn\comm
$$
Namely the commutator of $\Psi (l;|v\rangle )$ and 
$\Psi^\dagger (l';|v'\rangle )$ is nonzero only when $l=l'$ and $|v\rangle $ 
coincides with $|v'\rangle $ up to rotation. 
The string field Hamiltonian can be obtained from eq.\sdeq\ as 
$$
\eqalign{H=
&\sum_{|v_i \rangle }\int_0^\infty dl_1\int_0^\infty dl_2
\Psi^\dagger (l_1;|v_1 \rangle )\Psi^\dagger (l_2;|v_2\rangle )
\Psi (l_1+l_2;|v_1\rangle_{l_1}*|v_2\rangle_{l_2} )
\cr
&+g
\sum_{|v_i \rangle }\int_0^\infty dl_1\int_0^\infty dl_2
\Psi^\dagger (l_1+l_2;|v_1\rangle_{l_1}*|v_2\rangle_{l_2} )
\Psi (l_1;|v_1 \rangle )\Psi (l_2;|v_2\rangle )
\cr
&+\sum_{|v\rangle }\int_0^\infty dl
\Psi^\dagger (l;{\cal H}(0)|v\rangle )\Psi (l;|v\rangle )
\cr
&+\sum_{|v\rangle }\int_0^\infty dl\rho (l;|v\rangle )\Psi (l;|v\rangle ).}
\eqn\ham
$$
Here $\rho (l;|v\rangle )$ expresses the tadpole term and it has its support 
at $l=0$.

The string amplitudes can be expressed by using this Hamiltonian as follows:
$$
<w(l_1;|v_1\rangle )w(l_2;|v_2\rangle )\cdots w(l_n;|v_n\rangle )>
=
\lim_{D\rightarrow \infty }<0|e^{-DH}
\Psi^\dagger (l_1;|v_1\rangle )\cdots \Psi^\dagger (l_n;|v_n\rangle )|0>.
$$
The string field S-D equation can be obtained as 
$$
\lim_{D\rightarrow \infty }\partial_D<0|e^{-DH}
\Psi^\dagger (l_1;|v_1\rangle )\cdots \Psi^\dagger (l_n;|v_n\rangle )|0>
=0.
$$
It is obvious from the construction of $H$ that this S-D equation can be 
written as an integration of the S-D equation in eq.\sdeq. 

We can estimate the dimension of the geodesic distance $D$ from the above 
Hamiltonian. The scaling dimension of various quantities can be estimated 
most easily by considering terms involving strings on which all the spins are 
aligned. For example, for $c=1-\frac{6}{m(m+1)}$ string, the scaling dimension 
of $g$ is given as $[g]=L^{-\frac{2(2m+1)}{m}}$ which coincides with the 
matrix model result\refmark{\DS}. 
The dimension of $D$ becomes $[D]=L^{\frac{1}{m}}$. This fact may be 
checked by numerical simulations. 

Thus we have constructed the string field Hamiltonian using the three-Reggeon-
like vertices. We should however remark that eq.\ham\  is a formal expression. 
As was clear from the discussions in the previous sections, 
the states like $|1\rangle $ play important roles in the analysis of the 
S-D equations. However such states have divergent norms in the usual 
definition of the norms of states in CFT. Therefore we should adopt a 
different norm (e.g. one defined by Cardy\refmark{\cardy}) in 
eqs.\comm\ ,\ham\ to make the Hamiltonian applicable to such states. 
Accordingly the definition of the three-Reggeon-like vertices ought to be 
changed. We will pursue these problems elsewhere.

\chapter{Conclusions}
In this paper we proposed the continuum S-D equations for 
$c=1-\frac{6}{m(m+1)}$ string. It was checked that the S-D equations are 
consistent with all the known results of noncritical string theory. 
Especially the $W$ constraints were derived from the S-D equations. 
The $W$ constraints essentially come from the fact that the loop operator 
$w(l;({\cal H}(\sigma ))^{m-1}|1\rangle )$ is null. In the continuum picture, 
it was proved by using 
the results of boundary CFT. 

We constructed the temporal gauge 
string field Hamiltonian from the S-D equations. The Hamiltonian looks 
similar to the Hamiltonian of the light-cone gauge string field theory
\refmark{\SFT}, involving only three string interactions. 
Since the form of 
the Hamiltonian is almost the same for any $c$, it might be possible to 
construct the temporal gauge Hamiltonian in the same way 
for $c>1$ case, especially for the critical string. This will be left 
to the future investigations.

\ACK{
We would like to thank M. Fukuma, T. Kawai, Y. Kitazawa, Y. Matsuo, M. Ninomiya, 
J. Nishimura, N. Tsuda and 
T. Yukawa for useful discussions and comments. N.S. is supported by 
the Japanese Society for the Promotion of Science for Japanese Junior 
Scientists and the Grant-in-Aid for the Scientific Research from the Ministry 
of Education No. 06-3758.} 
\refout
\endpage
\figout
\bye